\documentclass{article}
\usepackage{graphicx} 
\usepackage{xcolor}  
\usepackage{lipsum}  
\usepackage{hyperref, cite, booktabs, inputenc}
\usepackage{float, longtable}
\usepackage{rotating, pdflscape}
\usepackage{afterpage, floatpag, lscape, caption, tcolorbox}
\usepackage{parskip}
\usepackage{amsfonts, placeins}
\usepackage{subcaption} 
\usepackage{geometry, array, ragged2e, amsmath, multirow, adjustbox, authblk}
\usepackage[T1]{fontenc} 
\usepackage{ulem}
\usepackage{amsfonts}  
\usepackage{wrapfig}

\title{Enabling Evolutionary Therapy in Metastatic Cancer Lacking Serum Biomarkers}
 
\author[1,2]{Eva Molnárová}
\author[3]{Ties A. Mulders}
\author[4]{Marcela Spee-Dropková}  
\author[1]{Louise M. Spekking} 
\author[1]{Sepinoud Azimi}  
\author[1]{Irene Grossmann} 
\author[2]{Anne-Marie C. Dingemans}  
\author[1]{Kateřina Staňková}

\affil[1]{Institute for Health Systems Science, Delft University of Technology, Faculty of Technology, Policy and Management, Delft, The Netherlands}

\affil[2]{Department of Pulmonary Medicine, Erasmus MC Cancer Institute, Erasmus Medical Center, Rotterdam, The Netherlands}

\affil[3]{Department of Radiology \& Nuclear Medicine, Erasmus Medical Center, Rotterdam, The Netherlands}

\affil[4]{Department of Radiology, Groene Hart Ziekenhuis, Gouda, The Netherlands}
\date{}

\newcounter{mybox}
\renewcommand{\themybox}{\arabic{mybox}}

\begin{document}
\maketitle

\begin{abstract}
Evolutionary therapy (ET) aims to steer tumor evolution by adjusting treatment timing and dosing to control rather than eradicate tumor burden. Clinical use requires reliable monitoring of tumor dynamics to inform mathematical models that guide therapy. In cancers such as metastatic castrate-resistant prostate cancer and relapsed platinum-sensitive ovarian cancer, ET models are informed by serial serum biomarkers. For cancers lacking reliable biomarkers, such as metastatic non-small cell lung cancer (NSCLC), radiographic imaging remains the primary method for treatment response assessment, typically using RECIST 1.1 criteria. RECIST, which tracks a few lesions with one-dimensional (1D) measurements and defines progression relative to the nadir, the smallest tumor burden recorded after treatment, was not designed to support ET. It may miss early regrowth, underrepresent tumor burden, and obscure disease trends. Using a virtual NSCLC patient model, we demonstrate that lesion selection and measurement dimensionality strongly affect progression detection. Two-dimensional metrics provide modest improvement, but only 3D volumetric measurements accurately capture both tumor burden and its dynamics --- key requirements for ET. To support ET in cancers lacking biomarkers, response assessment must evolve beyond RECIST by integrating volumetric imaging, automated segmentation, and potentially liquid biopsies, alongside redefining progression criteria to enable adaptive, patient-centered treatments.

\end{abstract}

\section{Introduction}

In metastatic cancer, standard of care treatments that are usually administered at the maximum tolerable dose (MTD) often induce a strong initial tumor burden decrease but fail to achieve durable control due to the evolution of resistance~\cite{Bray2024, Duan2023,Pienta2020, Keresztes2025,Stankova2019ResistanceGames}.

Evolutionary therapy (ET), also known as adaptive therapy, aims to steer tumor evolution to prevent or delay treatment-induced resistance~\cite{GatenbyBrown2020, West2023, Dujon2021, Stankova2019ResistanceGames, Viossat2021}. This paradigm is rooted in evolutionary game-theoretic models, which demonstrate that treatment shapes competition between sensitive and resistant cancer cell types~\cite{Zhang2022,Zhang2024, pressley_evolutionary_2021, Stankova2019, satouri2025stability, salvioli2024stackelberg, garjani2025evolutionary,Wolfl2022,kaznatcheev_fibroblasts_2019}.

Dose modulation protocols, such as the one used in the ACTOv trial in relapsed platinum-sensitive high-grade serous ovarian cancer (NCT05080556), aim to stabilize tumor burden rather than killing as many cancer cells as fast as possible~\cite{Mukherjee2024, Strobl2023, Cunningham2020}.
The ACTOv protocol adjusts treatment dosing based on changes in cancer antigen 125 (CA125) levels, reducing the dose if CA125 decreases by more than 25\% and increasing it if CA125 increases by more than 25\%. This approach creates personalized treatment cycles, allowing for flexible, tumor-specific dosing aimed at stabilizing the disease and minimizing side effects.
Zhang et al.'s evolutionary therapy protocol in metastatic castrate-resistant prostate cancer (NCT02415621) uses intermittent dosing, where treatment is paused once prostate specific antigen (PSA) drops to 50 \% of its baseline value and resumed when PSA returns to the baseline. This creates patient-specific treatment cycles that vary in duration depending on individual tumor dynamics~\cite{Zhang2017, Zhang2022, Salvioli2024}. These trials demonstrate the potential of ET when tumor burden can be frequently monitored through serum biomarkers, such as CA125 in the case of ovarian cancer and PSA in the case of prostate cancer~\cite{Cunningham2020, West2023, Salvioli2024}. 

Serum biomarkers enable frequent and minimally invasive monitoring, allowing treatment to be modulated based on individualized tumor burden dynamics rather than fixed schedules~\cite{Zhang2022, Mukherjee2024, Strobl2023, West2023, Fischer2015}. These protocols define progression relative to the initial tumor burden and/or emphasize stability over temporal maximal regression.

However, in cancers lacking reliable serum biomarkers, such as metastatic NSCLC, radiographic imaging remains the only feasible modality to track tumor burden. This reliance introduces challenges when moving towards ET, which requires frequent monitoring of treatment response~\cite{soboleva2025bringing}.
 Biomarker-guided ET protocols operate on principles fundamentally different from those that shape conventional imaging-based response evaluation.

The lack of reliable serum biomarkers in certain cancers presents a major challenge for adapting current imaging frameworks to support ET - a challenge this Perspective seeks to address.

\section{Why Is RECIST Incompatible with Evolutionary Therapy}
\label{section2}
One of the most widely used frameworks for imaging-based response evaluation in solid tumors, including NSCLC, is RECIST 1.1~\cite{RECISTGuidelines, Litiere2017, Seymour2017}. These criteria define target lesions, limit their number and rely on one-dimensional measurements (1D) to classify responses, assuming lesions are spherical. 

While widely adopted, RECIST was developed for cytotoxic therapies and does not account for the spatial and temporal heterogeneity of tumor evolution~\cite{Therasse2000, Aykan2020, Martens2014, Choi2005}. This can be problematic in cancers where individual metastases may evolve asynchronously and respond differently to treatment~\cite{Reck2024, Sullivan2020}.

RECIST 1.1 evaluates the response using the sum of the longest diameters (SLD) of up to five target lesions (no more than two per organ). Without the appearance of new lesions, progression is defined as a $\ge 20$ \% increase from the nadir (the smallest SLD recorded after treatment initiation), with an absolute increase of at least 5 mm.

RECIST has streamlined the assessment process, reducing radiologists' workload and offering standardized criteria for treatment evaluation~\cite{Therasse2000, RECISTGuidelines, Seymour2017, Litiere2017, Choi2005}. This reduction in workload is particularly evident when comparing RECIST to earlier methods such as the WHO criteria, which required 2D measurements and a greater number of target lesions to be tracked~\cite{WHO1979, Miller1981 }.  By shifting to 1D measurements and limiting the number of target lesions (from up to 10 in RECIST 1.0 to 5 in RECIST 1.1), RECIST significantly reduced the time and effort required for radiological assessments\cite{Therasse2000, RECISTGuidelines, Jaffe2010, vanpersijn2010recist}.

However, the RECIST 1.1 criteria present several challenges for ET, which requires continuous, trend-sensitive monitoring. For example, PSA- or CA125-guided protocols are based on the anticipated biomarker dynamics based on its frequent measurements, not relative to nadir~\cite{Zhang2022, Mukherjee2024}.
 These two protocols focus on maintaining stability or delaying resistance rather than achieving maximum regression, while multi-drug evolutionary therapies may aim at extinction as well~\cite{gatenby2020eradicating}. In the case of extinction therapy, the goal is to apply second-stike treatment at the point of lowest tumor burden to bring cancer under its extinction threshold. In such protocols, knowing the nadir and tumor burden distance from it might still serve a strategic purpose~\cite{gatenby2019first, gatenby2020eradicating}. Similarly, stabilization strategies might aim to maintain tumor burden at nadir levels. Therefore, some RECIST components could be repurposed for ET.

\section{Evaluating RECIST Limitations Through a Virtual Patient Model}
\begin{landscape}
\begin{figure}[ht!]
\centering
\includegraphics[width=1.2\textwidth]{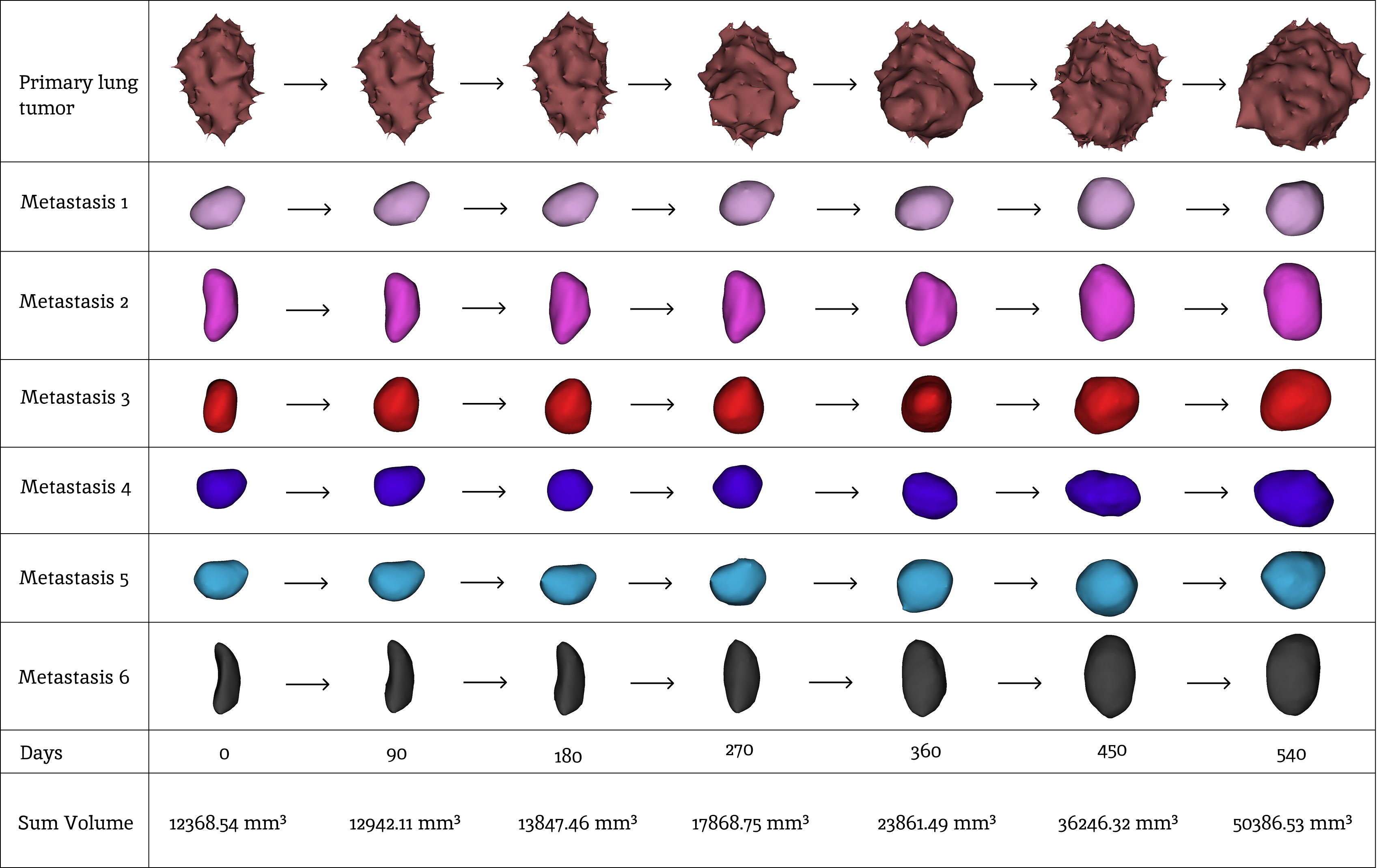}
\caption{Simulated progression of a virtual NSCLC patient with a primary lung tumor and six metastases over seven follow-up time points modeled at 90-day intervals. Using 3D Slicer, lesions were created to closely resemble the morphology and progression patterns of actual NSCLC tumors with validation by two radiologists. Each lesion was visualized in 3D and on standard clinical planes (axial, coronal, sagittal). While lesion images reflect realistic morphology, they do not represent actual anatomical scale. The total volume across all lesions increased from initial 12 368.54~mm$^3$   to 50 386.53 mm$^3$ on day 540, showing a continuous increase in tumor burden over time. These cumulative values are shown in the bottom row.}
\label{fig1}
\end{figure}
\end{landscape}

To explore whether components of the current imaging framework can be adapted to support ET in cancers without reliable serum biomarkers, we developed a virtual patient model representing metastatic NSCLC (Figure~\ref{fig1}). This model was constructed using data from the ongoing START-TKI trial (NCT05221372)~\cite{STARTTKI, Steendam2020, BROUNS2023, ernst2025utilizing, ernst2024hepatotoxicity, Bram2022, vanVeelen2023, veerman2023influence, DELEEUW2023, Pruis2023, DELEEUW2023weight, HEERSCHE2025, brouns2025connecting, ERNST2024, GULIKERS2024}, a multicentered observational trial with a large number of patient inclusions conducted in the Netherlands. 
This trial investigates the treatment of metastatic NSCLC with tyrosine kinase inhibitors, a type of targeted therapy, alongside other therapeutic agents, providing a robust dataset for modeling tumor progression and response to treatment. By using actual patient scans, we replicated the appearance and progression patterns of NSCLC lesions, ensuring biological and imaging realism which was confirmed by two clinical radiologists in our team. While RECIST 1.1 and ET have different cut-offs, some RECIST features, such as lesion selection rules and standardized measurement protocols, could be incorporated into ET strategies to maintain familiarity for radiologists transitioning between frameworks. Preserving these elements could also help maintain consistency and reproducibility in assessments. Furthermore, identifying the nadir may offer a practical target for extinction or stabilization therapies, as discussed in Section~\ref{section2}. Further details on the construction of the virtual patient model are provided in Box~\ref{box:virtualpatient}.

\refstepcounter{mybox}
\begin{tcolorbox}[title=Box \themybox\ | A virtual patient model to study lesion selection and measurement dimensionality]
\label{box:virtualpatient}

This virtual patient model simulates metastatic NSCLC progression in a controlled setting, allowing systematic investigation of how lesion selection and measurement dimensionality affect imaging-based response assessment. In addition to progression timing, we tracked tumor growth trends over time to evaluate how different measurement dimensions capture the tumor burden dynamics.

Lesions were generated in 3D Slicer~\cite{slicer}, an open-source software platform used for the visualization, segmentation and analysis of medical images. Each lesion was visualized in 3D and on standard clinical planes (axial, coronal, sagittal), mirroring clinical imaging practices. Seven simulated NSCLC lesions were created: one primary lung tumor and six metastatic sites in the brain, liver and adrenal glands shown in Figure~\ref{fig1}. These lesions were selected to reflect common, clinically measurable sites, frequently observed in the START-TKI trial. Lesion locations, growth patterns, and image segmentations were defined in collaboration with the two clinical radiologists to ensure biological plausibility and imaging realism. To simulate different lesion tracking strategies, we examined all combinations of 1 to 7 lesions and determined when each met progression criteria across different measurement dimensions.

Tumor burden was quantified using three methods: sum of longest diameters (1D, per RECIST), surface area (2D) and volume (3D), all derived from lesion segmentations. We defined nadir as the time point with the smallest sum of measurements in each dimension. Tumor evolution was modeled across seven follow-up time points at 90-day intervals, reflecting typical clinical imaging schedules.

\paragraph{1D – Diameter increase under RECIST 1.1:}

\[
D_{\text{new}} = 1.2D
\]

This equation reflects the RECIST 1.1 threshold for progression: a 20\% increase in the longest diameter ($D$) results in a new value $D_{\text{new}} = 1.2D$.

\paragraph{2D – Area scaling from a diameter increase:}

For 2D measurements, we approximate the lesion as a circle, where the area ($A$) is proportional to the square of the diameter ($D$):

\[
A = \frac{\pi}{4} D^2 \quad \Rightarrow \quad A_{\text{new}} = \frac{\pi}{4} (1.2D)^2 = \frac{\pi}{4} \cdot 1.44 D^2 = 1.44 A
\]

This shows that a 20\% increase in diameter results in a 44\% increase in surface area.

\paragraph{3D – Volume scaling from a diameter increase:}

For 3D measurements, we approximate the lesion as a sphere, where the volume ($V$) is proportional to the cube of the diameter:

\[
V = \frac{\pi}{6} D^3 \quad \Rightarrow \quad V_{\text{new}} = \frac{\pi}{6} (1.2D)^3 = \frac{\pi}{6} \cdot 1.728 D^3 = 1.728 V
\]

Thus, a 20\% increase in diameter corresponds to a 72.8\% increase in volume, which we round to 73\% for consistency.

\end{tcolorbox}

\section{One-Dimensional Measurements Capture Neither Tumor Burden nor Its Change Over Time}
One-dimensional measurements (1D), as utilized in RECIST 1.1, did not capture the actual time of disease progression. In our simulated patient model, the real time of progression was defined as the time at which the summed 3D volumetric burden across all seven measurable lesions crossed the progression threshold. Compared to this real time to progression, time to progression when using only 1D measurements was substantially delayed. This discrepancy was most pronounced when only one to three lesions were monitored (Figure~\ref{fig:combined_analysis}a–c), with delays of up to 270 days. Even when more lesions were included, 1D-based estimates remained variable and inconsistent with true disease progression, highlighting the sensitivity of the lesion selection.

To assess whether 1D-based progression detection systematically deviated from the real time of progression, we performed two-sided $t$-test for all lesion combinations that produced multiple progression observations. Significant differences were found for all 1D-based measurements when tracking 2 to 6 lesions ($p < 0.001$), with the largest discrepancies observed when 3 or 4 lesions were followed ($p < 10^{-7}$ and $p < 10^{-10}$, respectively). These statistical results underscore the inconsistency and delay introduced by one-dimensional tracking, even when multiple lesions are included.

In addition to the delayed progression detection, 1D measurements also underestimated the extent and pace of tumor growth. When the percentage change in total tumor burden from the nadir was tracked across seven follow-up time points, the 1D measurement curves were noticeably flatter than those based on 2D or 3D data. As a result, 1D measurements failed to capture key features of the growth trajectory, such as early signs of regrowth and periods of rapid progression, that were evident when using volumetric (3D) measurements.(Figure~\ref{fig:TumorBurdenTrends}).

These findings highlight critical limitations of RECIST 1.1 criteria. Restricting the number of lesions included in response evaluation can lead to missing the time to progression or estimating progression too late. 
Fundamentally, 1D tracking compresses complex, spatially heterogeneous tumor behavior into a single measurement, limiting our capacity to reflect real-time disease dynamics. In clinical contexts that require early recognition of progression or nuanced monitoring of tumor burden, such as within ET, the use of 1D measurements cannot be recommended, as it can obscure biologically significant changes in tumors and hinder timely adaptation of treatment.

\section{Two-Dimensional Measurements Capture Tumor Burden but Miss Temporal Change}

In our model, 2D surface area was derived from 3D segmentations and represents the total external surface of each lesion. When summed across all measurable lesions, it provided a consistent and anatomically detailed estimate of overall tumor burden at each timepoint, offering greater precision than simplified one-dimensional measurements.(Figure~\ref{fig:combined_analysis}).

Statistical analysis showed that 2D surface area measurements aligned more closely with the real time of progression than the 1D ones, in spite of  variation  with respect to the number of lesions. Significant differences were detected when tracking three and four lesions ($p = 0.009$ and $p = 0.005$), while comparisons at two and six lesions did not reach significance ($p = 0.17$ and $p = 0.36$). While the 2D measurements improve the tumor burden estimation when compared to the 1D ones, they capture the tumor burden progression dynamics across different lesion selections less closely.

When tumor burden was tracked longitudinally, 2D surface area measurements exhibited a steady rise from nadir, reflecting ongoing growth. However, as shown in Figure~\ref{fig:TumorBurdenTrends}, the 2D trajectory increased more slower than the total tumor burden did, especially after day 300, underestimating the rate of progression. The 2D values derived from the diameters further underestimated the tumor burden, illustrating the inaccuracy introduced by geometric approximations. Even when all lesions were included, both 2D curves showed dampened growth patterns compared to the total tumor volume.

These findings indicate that while the 2D surface area improves the tumor burden estimation at a given time point, it remains insufficiently sensitive to detect changes in growth dynamics over time. This makes the 2D surface area estimates less suitable for clinical settings that depend on estimation of how tumor burden changes over time, such as ET.

\section{Three-Dimensional Measurements Accurately Capture Tumor Burden and Temporal Change}

\begin{sidewaysfigure}
    \centering
    \includegraphics[width=1\textwidth]{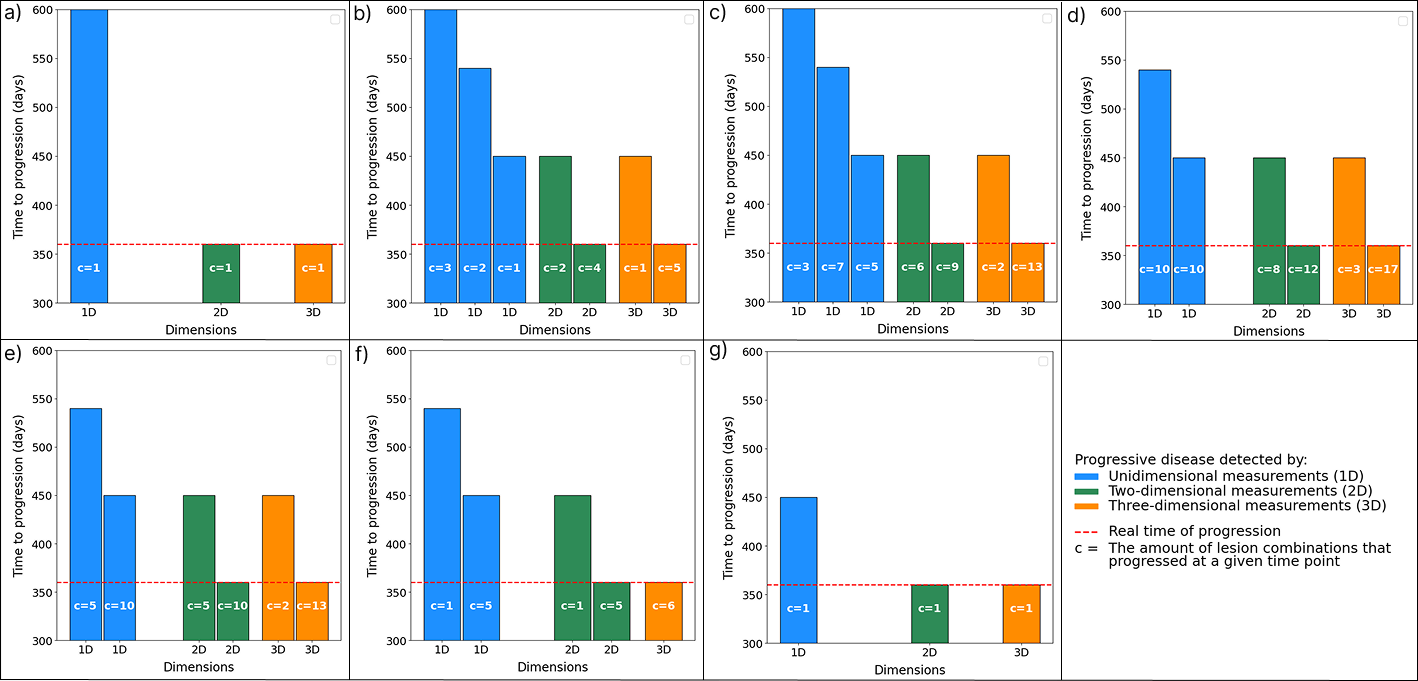}
    \caption{Progression timing was assessed using one-dimensional (1D; longest diameter), two-dimensional (2D; surface area) and three-dimensional (3D; volume) measurements. Panels (a–g) correspond to increasing numbers of tracked lesions: (a) one (primary only), (b) two (primary + one metastasis), (c) three, (d) four, (e) five, (f) six and (g) all seven lesions (primary + six metastases). For each lesion count, all possible combinations including the primary lesion were evaluated to simulate how lesion selection influences progression detection. The x-axis shows measurement dimensionality (1D, 2D, 3D), and the y-axis indicates time to progression (days), assuming 90-day imaging intervals. Each bar represents the number of lesion combinations that met progression criteria for the given dimensional method. The red dashed line marks the reference progression time (360 days), defined by progression detected using 3D volume measurements across all seven lesions. Bars reaching 600 days indicate combinations presumed to progress at the subsequent follow-up.}

    \label{fig:combined_analysis}
\end{sidewaysfigure}

As expected, the three-dimensional (3D) measurements provided the most comprehensive representation of tumor burden and its progression over time. In simulated progression detection (Figure~\ref{fig:combined_analysis}), 3D tracking consistently aligned with the real time of progression across all lesion counts, showing minimal variability and no significant delay, even when only a subset of lesions was monitored.

\begin{sidewaysfigure}
    \centering
    \includegraphics[width=0.95\textwidth]{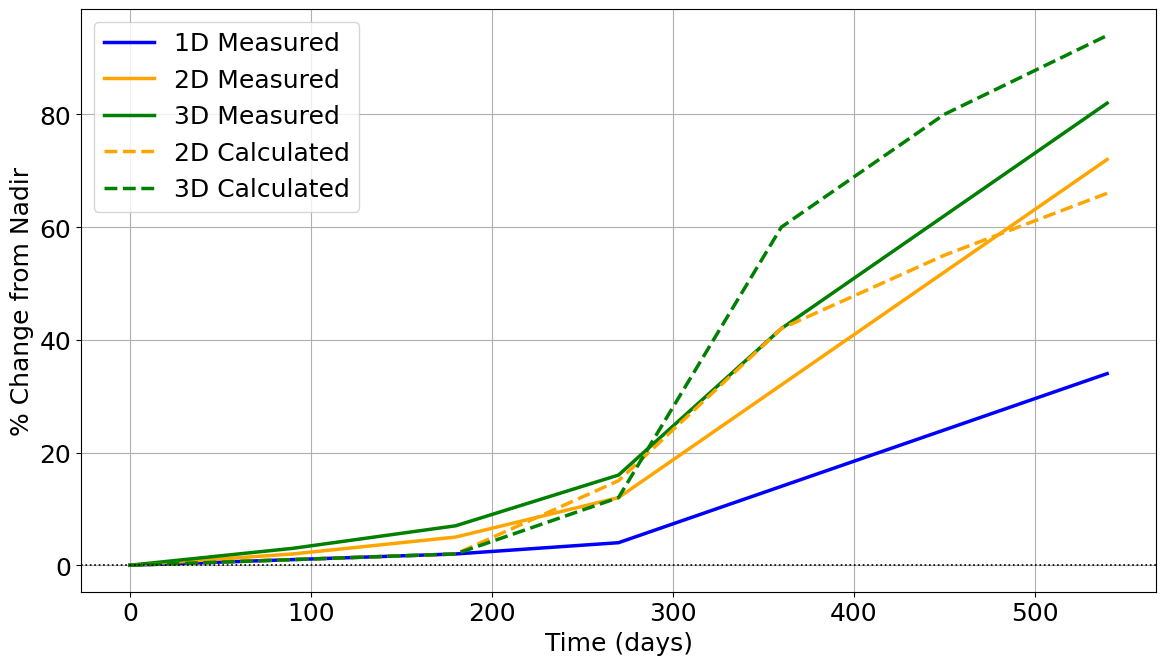}
    \caption{Mean percentage change in total tumor burden over time tracked from nadir using one-dimensional (1D: longest diameters), two-dimensional (2D: surface areas) and three-dimensional (3D: volumes) measurements. Solid lines represent direct measurements obtained from segmentation (1D: blue, 2D: orange, 3D: green), while dashed lines represent calculated estimates derived from 1D measurements using geometric approximations for 2D and 3D. Time (in days) is shown on the x-axis and percent change from nadir on the y-axis. Directly obtained 3D measurements show the greatest sensitivity to tumor growth over time, while 1D measurements exhibit the slowest rate of increase.}
    \label{fig:TumorBurdenTrends}
\end{sidewaysfigure}

When tumor burden was followed longitudinally (Figure~\ref{fig:TumorBurdenTrends}), the 3D measurements exhibited the most realistic trajectories, capturing early regrowth and periods of rapid expansion with greater precision than the lower-dimensional metrics. In contrast, 3D volumes calculated from thediameters underestimated the tumor burden and showed a slower rise, highlighting the limitations of geometric approximations.

No statistically significant differences were observed between measured 3D progression times and the reference at any lesion count (all $p > 0.3$), reinforcing the robustness of this approach even when lesion sampling is restricted.

By accurately capturing both the extent and pace of disease progression, directly measured 3D volume offers a reliable foundation for tumor monitoring. In the context of ET, where timely and trend-sensitive adjustments are critical, volumetric tracking is essential for guiding treatment effectively.

\newpage

\newpage
\clearpage
\pagebreak

\section{Conclusion}

Our findings highlight a fundamental mismatch between current clinical response criteria and the requirements of evolutionary therapy (ET). Designed to identify significant tumor shrinkage, RECIST 1.1 captures only a narrow view of tumor burden — tracking a few arbitrarily selected lesions in one dimension — and fails to reflect the spatial and temporal complexity of tumor evolution. This is particularly problematic for ET, which often aims not to eradicate the tumor but to stabilize its burden at levels higher than the nadir. Such strategies are incompatible with RECIST’s reliance on minimal tumor size as a benchmark for assessing progression. Even if we were to shift the reference point away from the nadir while keeping other elements of RECIST intact, the framework would remain inadequate for guiding ET.

A more holistic understanding of tumor dynamics calls for volumetric measurements. Yet, their routine adoption faces practical barriers: manual 3D segmentation is labor-intensive and costly, while heterogeneity in imaging protocols across centers complicates standardization. Although RECIST was originally developed to streamline radiologists’ workflow and improve reproducibility, it now risks constraining innovation at a time when more nuanced, dynamic assessment tools are urgently needed. Emerging AI-driven segmentation and burden-tracking tools hold promise to reverse this trend, offering faster, more reproducible volumetric assessments with reduced inter- and intra-observer variability~\cite{Bi2019, Dikici2020, Huynh2020, Chen2022, Ligero2025}.

Fortunately, these digital tools—ranging from automated 3D segmentation pipelines to cloud-based platforms for image processing—are rapidly becoming more scalable, reproducible, and clinically integrated~\cite{Ishikawa2024, Dahm2024, Aboian2022, rieke2020future, Echle2021}. Validated across multiple institutions, they enable real-time volumetric monitoring and are increasingly embedded in radiology workflows, closing the gap between clinical feasibility and research-grade analysis.

Complementing imaging, liquid biopsies — particularly analyses of circulating tumor DNA (ctDNA) — offer valuable insights into tumor heterogeneity, treatment resistance, and clonal dynamics. While challenges such as low ctDNA shedding and assay sensitivity still limit their application in NSCLC~\cite{leite2025plasma, Steendam2020}, recent advances underscore their potential for early detection of resistance and dynamic monitoring~\cite{Bertoli2023, Guibert2020, Tomasik2023, Rolfo2021}. Integrating volumetric imaging with ctDNA profiling could enable a multidimensional view of tumor evolution, bridging spatial and molecular dynamics.

Taken together, these innovations lay the groundwork for a shift from reactive to anticipatory cancer care. Realizing the full potential of evolutionary therapy will require not only new treatment strategies but also a reevaluation of how we measure and interpret tumor burden. Embracing more informative, scalable, and dynamic monitoring approaches is essential for delivering truly adaptive, patient-specific treatment.

\bibliographystyle{naturemag}
\bibliography{references}
\end{document}